\documentstyle[12pt]{article}

\input epsf

\newcommand{\be}{\begin{eqnarray}}
\newcommand{\ee}{\end{eqnarray}}

\begin{document} 
\title{New Possibilities for a Light Gluino}
\author{L. Clavelli\footnote{lclavell@bama.ua.edu}\\
Department of Physics and Astronomy\\ University of Alabama\\
Tuscaloosa AL 35487\\
} 
\maketitle
\begin{abstract}
Despite many positive indirect indications of light gluinos direct
searches for the expected signatures of gluino containing hadrons
have so far turned up negative severely restricting the allowable
windows in gluino mass.  After briefly reviewing the status,
we discuss a possible new decay scenario
that could have allowed light gluinos to evade direct detection 
with possible consequences for other measurements.
\end{abstract}
\renewcommand{\theequation}{\thesection.\arabic{equation}}
\renewcommand{\thesection}{\arabic{section}}
\section{\bf Introduction}
\setcounter{equation}{0}

   Recently several counter-indications to light gluinos have severely
eroded the attractiveness of the light gluino scenario. These are primarily

\noindent
 1.  New analyses of the running of the strong fine structure constant
     show consistency with standard QCD.\\
\noindent
 2.  Direct searches for gluino containing bound states at Fermilab have
     turned up negative (KTeV and E761).\\
\noindent
 3.  Concomitant predictions
     , in the Minimal Supersymmetric Standard Model (MSSM),
      of a light Higgs mass and a light chargino
     have been (at least marginally) ruled out at LEP II.

\noindent
Much theoretical and experimental effort could be spared if nature would
respect the current majority opinion that light gluinos are now excluded.
However, in the current paper we show how a modified light gluino scenario
might be viable in spite of these negative results.  There are still hints
from several experiments that provide motivation for further consideration
of light gluinos.   The discussion here is organized as follows.  In section
II we discuss the above counter-indications to a light gluino and some of
the proposed positive indications.  In section III we discuss a new
scenario for light gluino decay systematics, based on the idea of gauge
mediated supersymmetry breaking.  This scenario could loosen the constraints
from the negative direct searches.  In section IV we discuss some possible
experimental tests at LEP II and elsewhere.

\section{\bf Indications and Counter-Indications}
\setcounter{equation}{0}

   It has been known from the early days of Supersymmetry
(SUSY) to the present \cite{Fayet,Farrar} that a very light gluino
is a viable and theoretically attractive scenario.  In addition, in the
current decade, positive (though weak and indirect)
indications for such a light color-octet fermion have emerged
\cite{LC92,ENR,CCFHPY,Heb,CCY,JK,BB,Kagan,LC95,FarrarKolb,PFC,RS,CT1,CT2,CG}.
Although some of these require additional
(not unreasonable) assumptions such as flavor violating gluino
couplings or squarks of a particular mass, and some of the experimental
support for particular indications has eroded, sufficient reason remains
to further explore the light gluino option.

   Among the positive phenomenological indications of a light gluino that
have been noted are the following many of which are due at least partially
to the primary signature of a light gluino, an anomalously slow running
of the strong coupling constant.

    It is interesting to note that an earlier test
\cite{AEN} that reported negative results would have reached a positive
conclusion if current measurements of $\alpha_s(M_Z)$ had been available.
The QCD $\beta$ function is now known up to three-loop order including
gluino effects \cite{CCS1} so that the anomaly, if one exists, is very
unlikely to be due to higher order QCD effects.

    Analyses of data suggesting anomalously slow running have been done in
the quarkonium region \cite{CCY}, from the quarkonium to the Z region
\cite{LC92,JK,BM}, from deep-inelastic scattering \cite{BB}, and in
the Fermilab
jet inclusive transverse energy cross sections \cite{CT1,CT2}.  However,
some re-analyses of deep-inelastic data show consistency with the standard
model \cite{Alekhin,AK}.  In general, the lower one believes the value of
$\alpha_s$ is at low energies and the higher one believes its value is
at high energies, the more likely one is to be interested in the light
gluino option.  If one is willing to rely on QCD in $\tau$ decay and on
the current LEP measurements of $\alpha_s(M_Z)$, there is little room for
a light gluino.  However, the $\tau$ data is itself difficult to reconcile
with quarkonium and other low energy data.  One must either believe
that the apparent value of
$\alpha_s(M_\tau)$ is larger than the actual value due to some $10\%$
contribution from non-perturbative corrections or that the width of the
$J/\Psi$ is reduced by some $90 \%$ due to relativistic effects.

   If the $\tau$ data is misleading due to non-perturbative effects and the
$J/\Psi$ width gives a better estimate of the strong fine structure
constant, one can also understand the narrowness of the $\phi$
\cite{CCY} and then a light gluino is strongly favored.
Further discussion of the low energy measurements of $\alpha_s$ and the
"$\alpha_s$ problem" can be found in refs. \cite{CC,Shifman}.
More recent measurements of
the $R$ parameter at $LEP II$ seem not to require a slower running but
these are complicated by "radiative return" and W pair production and
they are not yet accurate enough to rule out a light gluino.
For instance, Delphi \cite{Delphi99} reports
\be
   \frac{d\alpha_s^{-1}}{d \ln E_{cm}} = 1.39 \pm 0.34({\it stat})
   \pm 0.17({\it syst})
\ee
to be compared with expected values 1.27 in the standard model and
0.95 in the light gluino case.  If the current LEP values of
$\alpha_s(M_Z) \sim 0.123
\pm .005$ and the low values from quarkonia analysis are both correct then,
not only is a light gluino strongly favored but, in addition, there must
be some additional effect tending to increase the apparent value of
$\alpha_s(M_Z)$.  It has been shown \cite{CCS2} that this effect could be
provided by virtual squark-gluino loops.

    Concomitant predictions of the light gluino, within the
gravity-mediated minimal
supersymmetric standard model, are those of a light higgs and a light
(${\cal O}(M_W)$) chargino and neutralino, together with low $\tan \beta$
near 2.  It could be considered encouraging that the indirect evidence
from LEP also favors a light higgs but direct searches now, at least
marginally, 
rule out these concomitant predictions.  The light Higgs is now
apparently above $90 GeV$ and, in the light gluino scenario, the
lightest chargino is experimentally at least $55 GeV$ \cite{Opal97} to
be compared with a maximum prediction of about $70 GeV$ for both the
lightest chargino and the light Higgs in the MSSM light gluino case.
Thus the light gluino is ruled out 
in the MSSM
unless there are
perturbations to these mass predictions or gaugino mass
universality is broken.

    Most recently, the CDF data on jet inclusive transverse energy cross
sections and on the scaling ratio \cite{scaling,Bhatti}
at two different beam energies have
been shown to be in line with light gluino expectations.  In addition,
structure in the scaling function has the right features (dip then bump
separated by a factor of $1.8/0.63$ in $X_T$, with approximately the
right height and width) to suggest the existence of valence squarks in
the 106 to 133 GeV mass range \cite{CT2,run2}.  In the light gluino case 
such a
relatively low mass squark would have evaded the Fermilab searches since
it would decay into quark gluino without the standard high transverse
energy lepton and missing energy signal.  It should be noted, however,
that $D0$ data, while consistent with a light gluino, do not confirm 
structure in the scaling ratio which would indicate the presence of a
valence squark \cite{D0}.

     In contrast to these possible indirect hints of the existence of a
light gluino, all direct searches have so far turned up negative.
These have been based on the standard expectation that the light gluino
would decay to quark-antiquark-photino through an intermediate squark
with the approximate lifetime
\hfill\break\vfill
\be
 \tau_{\tilde g} \sim \frac{m_{\tilde Q}^4}{\alpha \alpha_s{m_{\tilde g}}^5}
 .
\label{gamma1}
\ee
\begin{figure}
\centerline{\epsfxsize=4in \epsfysize=4in \epsfbox{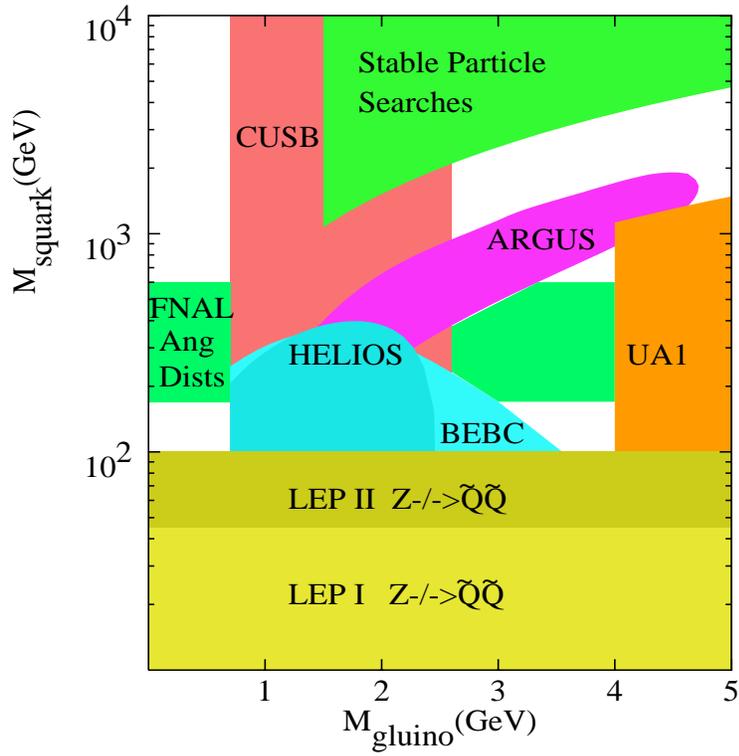}}
\caption{Current map of gluino windows.  The regions ruled out by
various considerations
\protect\cite{stable,HRD,Terekhov,cusb,argus,ua1,helios,bebc}
are indicated.  Some of these such as the beam dump constraints
and stable particle searches are dependent on the SUGRA-inspired
relation between gluino lifetime and squark mass.} 
\end{figure}

A light gluino would be expected to hadronize into a gluino-gluino state
(gluinoball), a gluino-gluon
state (glueballino, $R^0$), a quark-antiquark-gluino state (mesino),
or a three quark + gluino state (barino).  The latter two types of
gluino bound states would include new charged hadronic states
for which there are stringent experimental limits.  The gluinoball would
decay rapidly and be
virtually indistinguishable from a glueball state.  There are in fact
too many good glueball candidates to be explained by gluon composites
only.  However, there seems to be no reason why a large fraction of
produced gluinos should not hadronize into glueballinos which do have
a distinctive signature.  One could theorize that mesinos and barinos
are not formed for the same reason that there are no candidates for
hybrid gluon containing states ($qqqg$ or $q{\overline q}g$):  The
QCD potential is repulsive in the requisite color octet $qqq$ or
$q {\overline q}$ sub-state.

    With the squark near or above the Z in mass the glueballino would
be long-lived, depositing most of its energy in a hadronic
calorimeter before decaying, and leaving only a small amount of energy
to the missing photino.  A produced gluino hadronizing into a
glueballino would behave, therefore, very similarly to a normal gluon jet.
This signature has been extensively reviewed
\cite{LC92,Farrar95} and would allow a light gluino to avoid the missing
energy trigger that puts a high lower limit on the mass of
conventionally decaying SUSY partners.

    However, the long lifetime of the glueballino makes it sensitive to
other search techniques.  The most stringent restriction on the $R^0$
state comes from the KTeV experiment \cite{KTeV} which
convincingly rules out a long
lived neutral particle 
with a lifetime up to $10^-8 s$
contaminating a $K_L$ beam and decaying into
$\pi^+ \pi^- \em +$ missing energy with the $\pi \pi$ state having
joint invariant mass above the Kaon mass.  In its current analysis
KTeV requires such a high $\pi \pi$ mass to avoid background from
CP violating $K_L$ decays. One way out would be to
have the photino mass so close to the glueballino mass that such a
high mass hadronic final state would be kinematically excluded
\cite{Farrar97}.  It seems likely that KTeV can also explore this
possibility with later refinements in the analysis.  A second way
out would be to have the glueballino lifetime so long that it does
not decay in the KTeV sensitive region.  
If such a glueballino were not much heavier than the nucleon, its decay 
could be confused with neutron inelastic collisions
which limit the bounds set by 
stable particle searches such as \cite{stable}.  Recently various
authors have even considered an absolutely stable gluino \cite{Raby,BCG}.
The consequent stable $R^0$ would have been
evident in missing energy searches if its mass were greater than 
about $5 GeV$ \cite{BCG}.  Below about $2 GeV$ a stable $R^0$ cannot at
present be conclusively ruled out since the signature knock-on events
would not be distinguishable from neutron induced events for which the
cross sections are poorly known.  The viability of a stable
$R^0$, of course, depends on such a particle not being able to bind
to nuclei to create anomalous isotopes.  The absence of such bound
states might be understandable due to the inability of the flavor singlet
glueballino to interact with nuclei via pion exchange (or even 
more massive meson exchange).  The dominant interaction mechanism
would be glueball
exchange which is presumably too weak to lead to a thermally stable
bound state.  The same could not be said in the case of a stable mesino
or barino which would then lead to a great number of exotic anomalous
isotopes. Such particles, therefore, are probably not phenomenologically
viable.  Further direct counter-indications come from the $E761$ experiment
\cite{E761} which, however, would be voided if charged gluino containing
hadrons are not bound due to the argument suggested above.

   Perhaps the most constraining recent evidence concerning light gluinos
comes from the dijet angular distributions measured at Fermilab and
interpreted in terms of a squark decaying into a quark plus a light gluino
\cite{HRD,Terekhov}.  The absence of deviations from the standard model
rules out, in the light gluino case, a valence squark between $150$ and
$650 GeV$.  In Tevatron run II it should be easily possible to extend this
range above $1 TeV$.  If the predictions for scaling violations between the
$2 TeV$ and $1.8 TeV$ data are not borne out \cite{run2}
the window between $100 GeV$ and $150 GeV$ will also
be closed.  This would make the light gluino scenario very unattractive
since squarks above a $TeV$ would lead to extreme fine tuning problems for
supersymmetry.  In summary the constraints on gluino mass
\cite{stable,HRD,Terekhov,cusb,argus,ua1,helios,bebc}
in the low energy windows in the current standard
picture of gluino decay systematics are shown in Fig. 1.
Earlier versions
of this map can be found in references \cite{DEQ,ua1}.  In Fig. 1
we have updated the plot to include recent 
LEP and Fermilab constraints.
The sloping lines in the constraints are based on the
assumption that the gluino lifetime would be that of the Sugra-based
model.  In the next section we consider a gauge mediated model which would
drastically change the relation between the gluino lifetime and squark
mass.
  
\section{\bf Gauge Mediated Model}
\setcounter{equation}{0}

     We turn now to a discussion of a new light gluino scenario, guided by
the gauge mediated supersymmetry breaking (GMSB) ideas, which could avoid the
problems presented by the negative direct searches while preserving
the positive indications.  It has been noted that there is an attractive
scenario within the GMSB scheme in which the gluino is the lightest
supersymmetric particle (LSP) or the next-to-lightest with an ultra-light
gravitino being the LSP \cite{Raby}.  There is no natural lower limit in
these scenarios to the gluino mass and, in the low energy window its
mass must, in fact, be below about $5$ GeV \cite{BCG}.  In the GMSB
models, the gluino mass is often naturally decoupled from the chargino
and neutralino masses so that the constraints on the latter
are easily avoided.  Similarly, once the MSSM predictions on chargino
mass are eliminated, the light gluino constraint on $\tan \beta$ is
relaxed allowing for somewhat higher (but still low enough to be
phenomenologically interesting) predictions for the Higgs mass.  The
experimental bounds on the Higgs mass are, in fact, also somewhat relaxed
by the Higgs decay to two gluinos.
The same results can also be obtained in any model giving up
gaugino mass universality such as that considered by \cite{CCOPW}.
Consistent with this model, we tentatively propose that the light Higgs
and gaugino masses are in the $100 GeV$ region except for the gluino
which is below (perhaps far below) $5 GeV$.  If the $R^0$ is absolutely
stable, providing it does not bind to ordinary nuclei to create
anomalous isotopes it is then invisible to current direct searches.

    We would, however, like to consider the alternative in which the
gluino decays into an ultra-light gravitino and a gluon, the standard
GMSB decay mode.  The gravitino couples to the supercurrent with a
strength, $F$,
\be
 {\cal L} = - \frac{1}{F} j^{\alpha,\mu} \partial_{\mu} G_{\alpha}
       + h.c.
\ee
Thus the gravitino couples the NLSP to its partner, in this case the gluon.
F is related to the squark and slepton masses by
\be
        F \approx \big {(4 \pi/\alpha)^2} m_{\tilde Q}^2
\ee
where $\alpha$ is the mediating gauge coupling constant.
Assuming that $\alpha$ is between the ordinary fine structure constant
and unity,
\be
       F \approx (10^2 \em to \em 10^6) m_{\tilde Q}^2    .
\ee

The gluino decay width is then
\be
      \Gamma \approx m_{gluino}^5/F^2 \approx m_{gluino}(m_{gluino}/m_{squark})^4
            \times (10^{-4} \em to \em  10^{-12})
\ee
or, taking nominal values of 130 GeV and 130 MeV for
the squark and gluino masses respectively,
\be
      \tau(gluino) \approx (10^{-7}  \em to \em  10) s  .
\ee

In fact, since
the gravitino couples particles to their superpartners and since the gluino
lifetime is much longer than the hadronization time, the relevant decay is
\be
      R^0 \rightarrow f^0 + {\tilde G}  .
\ee
Here $f^0$ denotes the glueball partner of $R^0$.  Since with a light gluino
we have an approximate supersymmetry in the gauge sector, $f^0$ and $R^0$
are expected to be approximately degenerate with $R^0$ only slightly
heavier.  Consequently the gravitino does not carry off enough energy to
be caught by a missing energy trigger.
In addition, the multi-pion decay of the glueball, coupled with a
lifetime of up to $10 s$ would make the state invisible to
the current phase of KTev.

\section{\bf Conclusions}
\setcounter{equation}{0}

    In the suggested scenario, all produced susy particles would decay
strongly down to the $R^0$ which would then decay with a long lifetime
and no appreciable missing
energy to the lightest glueball.  
The UA1, Bebc, Helios, and stable particle search constraints in Figure 1
are then no longer operative.  The CUSB constraint depends on a model for
the gluinoball wave function at the origin and should perhaps not be
regarded as a strict exclusion.
The primary
signature of SUSY production 
(e.g. charginos)
at LEP II would be an excess in the visible
energy cross section since standard model background would have
appreciable energy loss to neutrinos from charm and bottom decays.  Such
an excess above the standard model monte carlos can in fact be
perhaps seen in
Figure 1 of the L3 note \cite{L32227}.  The light gluino predictions for
the scaling violations at Fermilab would be preserved by this new decay
scenario.

    In the current model for gluino decay, one might also expect excess
glueball production in Upsilon and B decays
due to final states containing a gluino pair
\cite{CCY,PFC}.  This could be related to
the observed anomalous $\eta \prime$ production in B decay \cite{CLEO}
assuming the $\eta \prime$ has a significant glueball component.

    In summary we feel that neither the weak hints in favor of a light
gluino nor the counter-indications can be considered conclusive at this
time and further experimental tests are needed.  Further analysis of the
LEP II data and the Fermilab Run II data along the lines we have
suggested could provide the crucial tests for the existence of low
mass SUSY particles.

    This work was supported in part by the US Department of Energy under
grant no. DE-FG02-96ER-40967.


\end{document}